\pdfoutput=1

\RequirePackage[hyphens]{url}
\documentclass[a4paper,11pt]{article}
\usepackage{jheppub}
\usepackage[section]{placeins}
\usepackage{bm}
\usepackage{appendix}
\usepackage[hyphens]{url}
\usepackage{hyperref}
\usepackage{amsfonts}

\usepackage{float}
\usepackage{subfig}
\usepackage{cancel}
\usepackage{lscape}

\usepackage{color}
\usepackage[usenames,dvipsnames,svgnames,table]{xcolor}

\usepackage{amsmath}
\usepackage{amssymb}
\usepackage{graphicx}
\usepackage{slashed}
\usepackage{soul}
\usepackage{multirow}

\def\lsim{\mathrel{\raise.3ex\hbox{$<$\kern-.75em\lower1ex\hbox{$\sim$}}}}
\def\gsim{\mathrel{\raise.3ex\hbox{$>$\kern-.75em\lower1ex\hbox{$\sim$}}}}

\newcommand{\bea}{\begin{eqnarray}}
\newcommand{\eea}{\end{eqnarray}}

\newcommand{\be}{\begin{equation}}
\newcommand{\ee}{\end{equation}}

\title{Personal Ultraviolet Respiratory
 Germ Eliminating Machine~(PUR$\diamond$GEM) for COVID-19: \\
 Prototype Development}
\author[a]{Nausheen~R.~Shah,}

\affiliation[a]{Department of Physics \& Astronomy, Wayne State University, Detroit, MI 48201, USA}

\emailAdd{nausheen.shah@wayne.edu}

\preprint{WSU-PHY-2002}

\abstract{
In this article we detail the prototype development of the Personal Ultraviolet Respiratory
 Germ Eliminating Machine~(PUR$\diamond$GEM) to safely, efficiently and economically, continuously disinfect inhaled/exhaled air using Ultraviolet~(UV) radiation with possible 99.99\% virus elimination. The PUR$\diamond$GEM consists of a series of UV disinfection chambers through which constant airflow is maintained via fans. A minimum air flow rate of $\sim20-30$ L/min is sufficient to keep CO$_2$ levels $\lesssim$ 0.5\% in the hood/helmet. We  validated that using easily available PTFE wrap, a factor of $\sim 18$ enhancement in UV power can be obtained in our spherical chambers. Detailed analysis is presented for the air travel time distributions through the cavities, and the expectation value for actual pathogen elimination is computed. We provide the scaling of pathogen elimination with the number of cavities in series, reflective enhancement, UV source power, sphere radius and airflow rate. We show that disinfection greater than 4-log is achievable for a series of three or more spheres.  We 3D printed our prototype, consisting of two spheres of 10~cm diameter each in series for each direction of disinfection. Using UVC LEDs emitting $\sim$ 40~mW of power each with an airflow rate of 30 L/min, actual SARS-COV2 virus elimination of $\sim$ 98\% is expected. While not manually feasible to construct smaller spheres in the lab, smaller cavities can be commercially manufactured, leading to significantly higher actual pathogen elimination, as well as reducing fingerprint and cost of cavity manufacture. Patent pending. 
}

\begin{document}

\maketitle

%**********************************************************************
\section{Introduction} \label{sec:intro}
%********************************************************

The Personal Ultraviolet Respiratory Germ Eliminating Machine~(PUR$\diamond$GEM) was proposed in Ref.~\cite{PURGEM} as a reusable wearable portable device to safely disinfect inhaled/exhaled air. A highly reflective spherical  cavity was proposed to enhance initial Ultraviolet~(UV) source power irradiating inflowing air. As an example material, Polytetrafluoroethylene (PTFE)~\cite{PTFE}  is a widely and cheaply available UV resistant material with $> 95\%$ reflectance~\cite{weidner1985laboratory}. Due to the high reflectance and spherical shape, the irradiance is expected to be uniform throughout the sphere.  
Further, a series of cavities is proposed to obtain efficient actual disinfection. 

In this article, the prototype development of the PUR$\diamond$GEM is detailed. In Sec.~\ref{sec:UVC} the results from  investigations on the UVC LED power, reflectance of several materials and consequent UV power amplification in spherical cavities are presented. It was found that multiple layers of thin PTFE wrap provided a factor $\sim$18 amplification and was used in the prototype constructed. 
Sec.~\ref{sec:series} presents the analysis for the air travel time distributions through the cavities and the computation for the  expectation value for actual pathogen elimination. The scaling of pathogen elimination with the number of cavities in series, reflective enhancement, UV source power, sphere radius and airflow rate is provided. It is shown that disinfection greater than 4-log is achievable for a series of three or more spheres. Sec.~\ref{sec:proto} presents the prototype built together with a simple hood configured to couple with the PUR$\diamond$GEM to make a complete practically functioning device.  The prototype was mostly 3D printed, consisting of two spheres of 10~cm diameter each in series for each direction of disinfection. Each spherical cavity was layered with 4 layers of PTFE wrap. Using UVC LEDs emitting~(measured) $\sim$ 40~mW of UVC power each with an airflow rate of 30 L/min, actual virus elimination of $\sim$ 98\% is expected. While not manually feasible to construct smaller spheres in the lab, smaller cavities can be commercially manufactured, leading to significantly higher actual virus elimination, as well as reducing fingerprint and cost of cavity manufacture. 

%**********************************************************************
\section{UV Power, Reflectance and  Amplification} \label{sec:UVC}
%**********************************************************************

UVC LEDs with a peak wavelength of 265 nm, almost perfect Lambertian emission, and purported to be 70 mW were obtained from Klaran~\cite{LEDs}. A data sheet containing the spectral intensity, thermal derailment etc. was provided by the manufacturer. A UVC 254nm Ultra Violet Light Meter (UVC-254SD)~\cite{meter} was used to measure UVC power. The responsivity curve for the meter, obtained from the manufacturer, showed good sensitivity for the spectral output of the LED. Convoluting the sensor responsivity curve to the spectral output of the LED corresponds to $\sim 93\%$ accuracy in measuring the LED output. 

 \begin{figure}[tb!]
   \begin{centering}
       \includegraphics[width = \textwidth, trim = 3cm 15cm 3cm 2cm, clip]{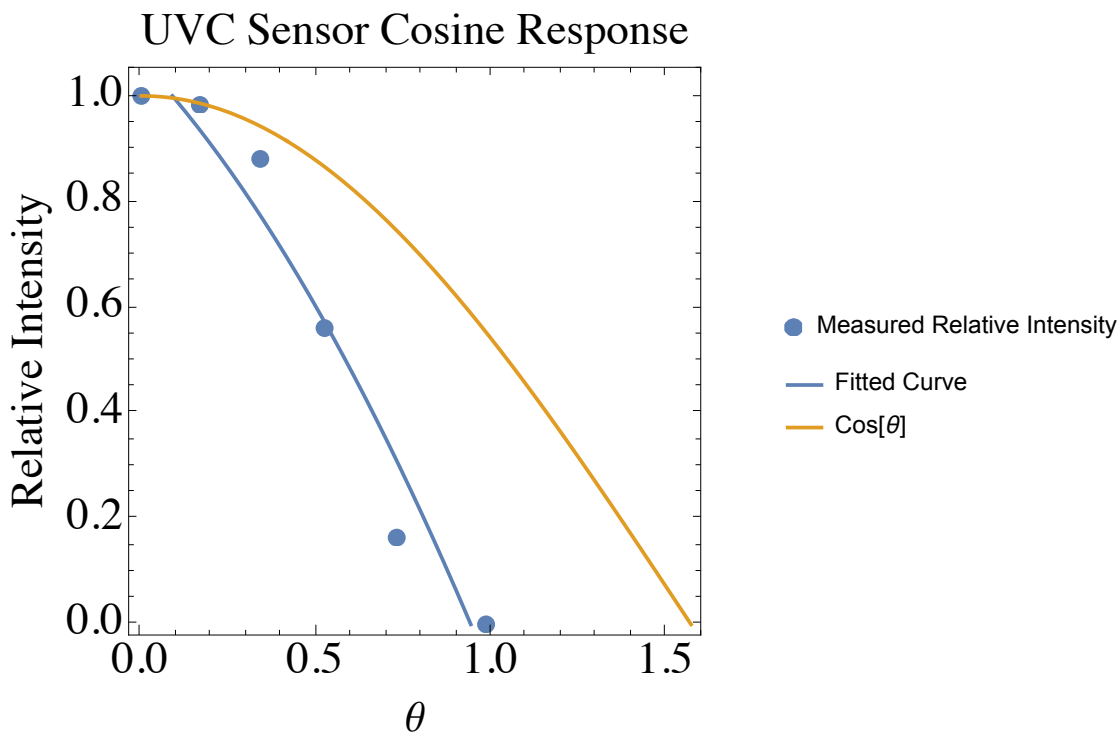}
      \caption{Relative intensity as a function of the angle of incidence is shown. The blue points denote the average value measured at a distance of $d=15$ cm between LED and UV sensor where the sensor is positioned at an angle $\theta$ with respect to the LED. The blue line denotes the fitted curve whereas orange curve shows the perfect cosine response.  See text for details. }
      \label{fig:cos}
   \end{centering}
\end{figure}

The sensor face was covered with a small buffer, such that only a 1 cm diameter centered region of the sensor face was exposed. This was done  to correspond to situation when making measurements for the UVC power inside the spherical cavities through a circular port. The cosine response of the sensor was then measured experimentally. The sensor was placed at distances $d=10, 13$, and 15 cm away from the LED respectively, and measured power was recorded as the angle of incidence was changed. Each measurement was repeated 5 times, and consistent results were found at each distance. The measured average values for $d=15$ cm are shown in Fig.~\ref{fig:cos} scaled to the value at $\theta=0$ as blue points. The blue line shows a fitted curve for the measured points and the yellow curve shows the perfect cosine response. As can be seen, while at small angles, the sensor follows the perfect cosine curve, the sensitivity falls  sharply for angles $\theta \gtrsim 0.3$, and there is effectively no sensitivity if the angle of incidence is $\gtrsim 1$.  Integrating the fitted curve and comparing to the cosine curve, it was found that the power measured by the UVC sensor is $\sim$ 30\% of the all angle uniform incident radiation power.

The UVC meter was then used to measure the power output of the LEDs by making power measurements as a function of the distance for $d$ varying from 10 - 30 cm. Under the assumption of a Lambertian emitter~($P_{total}=\pi d^2$), it was consistently found that for all the LEDs used in constructing the prototype, these measurements corresponded to LED power $\sim$ 40 mW, significantly lower than the listed 70 mW, even accounting for the approximately 10\% thermal derailment expected.  The LED manufacturer was contacted but an explanation for the discrepancy has not been found. Therefore, for the purposes of disinfection calculations for the prototype, a value of 40 mW will be used for source power. The almost perfect Lambertian nature of the LEDs was verified by making relative angular measurements around the LED, taking into account the cosine response of the sensor.   

\begin{figure}[tb!]
   \begin{centering}
       \includegraphics[width = 0.4\textwidth, trim = 5cm 19cm 10.5cm 4cm, clip]{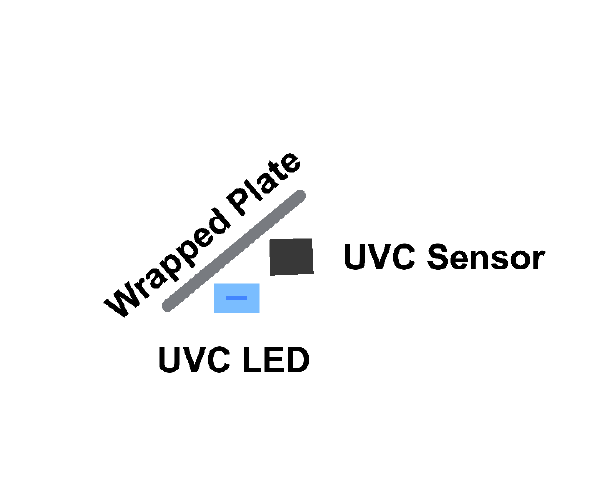}
       \hspace{.5in}
       \includegraphics[width = 0.4\textwidth, trim = 5cm 19cm 10.5cm 4cm, clip]{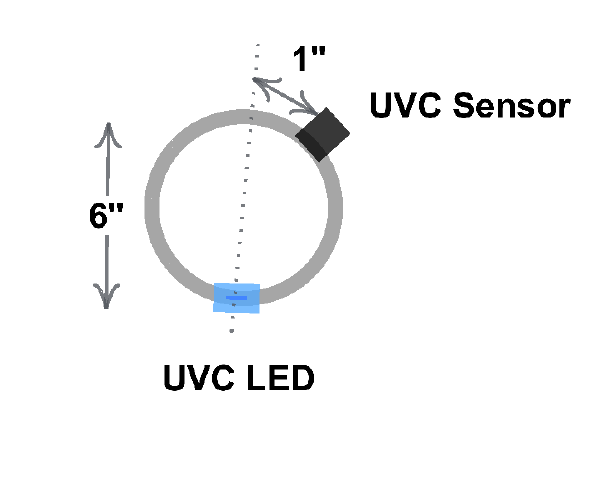}
      \caption{ {\it Left:} A schematic for the set-up used for measuring UV reflectance of various materials. A plate wrapped with the material being tested  was placed at 45$^\circ$ to both the sensor and the LED, and power measurements were recorded. {\it Left:}  Drawing not to scale. See text for details. }
      \label{fig:Exp}
   \end{centering}
\end{figure}

Relative reflectance of various materials was tested.  A schematic of the set-up used  is shown in the left panel of Fig.~\ref{fig:Exp}. 
Each material was wrapped around a flat plate which was  placed at 45$^\circ$ to both the sensor and the LED. Power measurements were then recorded. 
Materials included household aluminum foil, an assortment of common PTFE/Teflon tapes found at hardware stores, as well as special ordered PTFE rolls and films. The highest relative reflectance was found for PTFE ``relic wrap"\cite{relic}. Since PTFE is a depth reflector, reflectivity for multiple wrappings was measured. It was found that $>$ 4 layers did not increase the reflectance further.

Due to the reflectance of the walls in a sphere, one gets a geometric enhancement factor for the UV power. This is reduced by any defects in the reflecting surface, such as those due to the surface area occupied by the LED and the air vents. Therefore, the effective reflectance $R_{eff}$ of the cavity is given by:
 \be
 R_{eff}= R\left(\frac{SA_{s}-A_{def}}{SA_{s}} \right)\; ,
 \ee  
 where $R$ is the reflectance of the walls, $SA_{s}=4\pi r^2$ is the surface area of a sphere of radius $r$,  and $A_{def}$ is the total area of defects.  The enhancement due to effectively infinite reflections of light inside the cavity  gives a multiplicative factor to $P_{UV}$:
 \be
 \mathcal{E}=\frac{1}{1-R_{eff}}\;. \label{eq:ref}
 \ee

To measure this power enhancement in a spherical cavity, cheaply available 6" acrylic hemispheres were used, where two hemispheres were then clamped together at the equator to form a sphere. A 2 cm hole was drilled at~(slightly off-center) bottom of each sphere for the LED port, such that the sphere is slightly tilted to the side when placed on top of the LED.  Each sphere further had a 1 cm hole drilled on the opposite side approximately 1" off center used as the viewing port for the UVC sensor. A schematic is shown in the right panel of Fig.~\ref{fig:Exp}. Since acrylic completely absorbs UVC in the range of wavelengths we are investigating, a clear sphere was used as the control for source power. Layers of PTFE relic wrap was applied to the test spheres using Elmer's glue, and power measurements were taken at the sensor port, always comparing with free standing LED at a certain distance and in the clear sphere to confirm unreflected LED power. Amplification was computed using the excess power measured in test spheres as compared to clear sphere, taking into account the cosine response of the sensor.  Further, power measurements were recorded at different angles as the sphere was tilted around on top of the LED so that the incident radiation would fall on different surface defects at different distances. Measurements were also taken with the sensor port and LED port switched. Excellent agreement was found at different angles and orientations, validating diffuse uniform radiation inside the sphere. The sensor was also placed inside the test spheres at different locations and measurements recorded. They were roughly consistent with even radiation, but due to the large sensor size contributing to the surface defects in reflectivity, it was difficult to be quantitive.  While PTFE reflectance is somewhat lower at visible wavelengths, it was  visually verified that it was almost impossible to see any defects inside the test sphere, again validating our expectation of uniform radiation. The amplification in the visible band was even more striking in the spheres used for the prototype which were 3D printed from black PLA. Before lining the interior with PTFE relic wrap, it was very difficult to see anything inside the sphere, even when a light was shone inside. However, after the PTFE lining was applied, a very even, bright glow could be observed in the interior from the LED port~(with ambient outside light), and only the edges of the air ports were barely perceptible.  

Quantitatively it was found, unlike for the plate, that there was no difference in amplification between 3 and 4 layers. It was realized that while wrapping the plate could be done almost perfectly smoothly, when lining the interior of the hemispheres, there were multiple areas where overlapping occurred. Hence, even at 3 wrappings, there was a significant fraction of the surface area which already had 4 or more layers. An amplification of $\sim 18$ was obtained for both 3 and 4 wrappings for the PTFE relic wrap, which corresponds to 94.4\% reflectivity according to Eq.~\ref{eq:ref}. Measurements were also made with linings of other materials, and were found to be consistent with previously  reported UVC reflectivity of such materials~\cite{6168236}.

 %**********************************************************************
\section{Series Cavity Disinfection} \label{sec:series}
%**********************************************************************

 \begin{figure}[hptb!]
   \begin{centering}
       \includegraphics[width = \textwidth, trim = 0cm .7cm 0.5cm .5cm, clip]{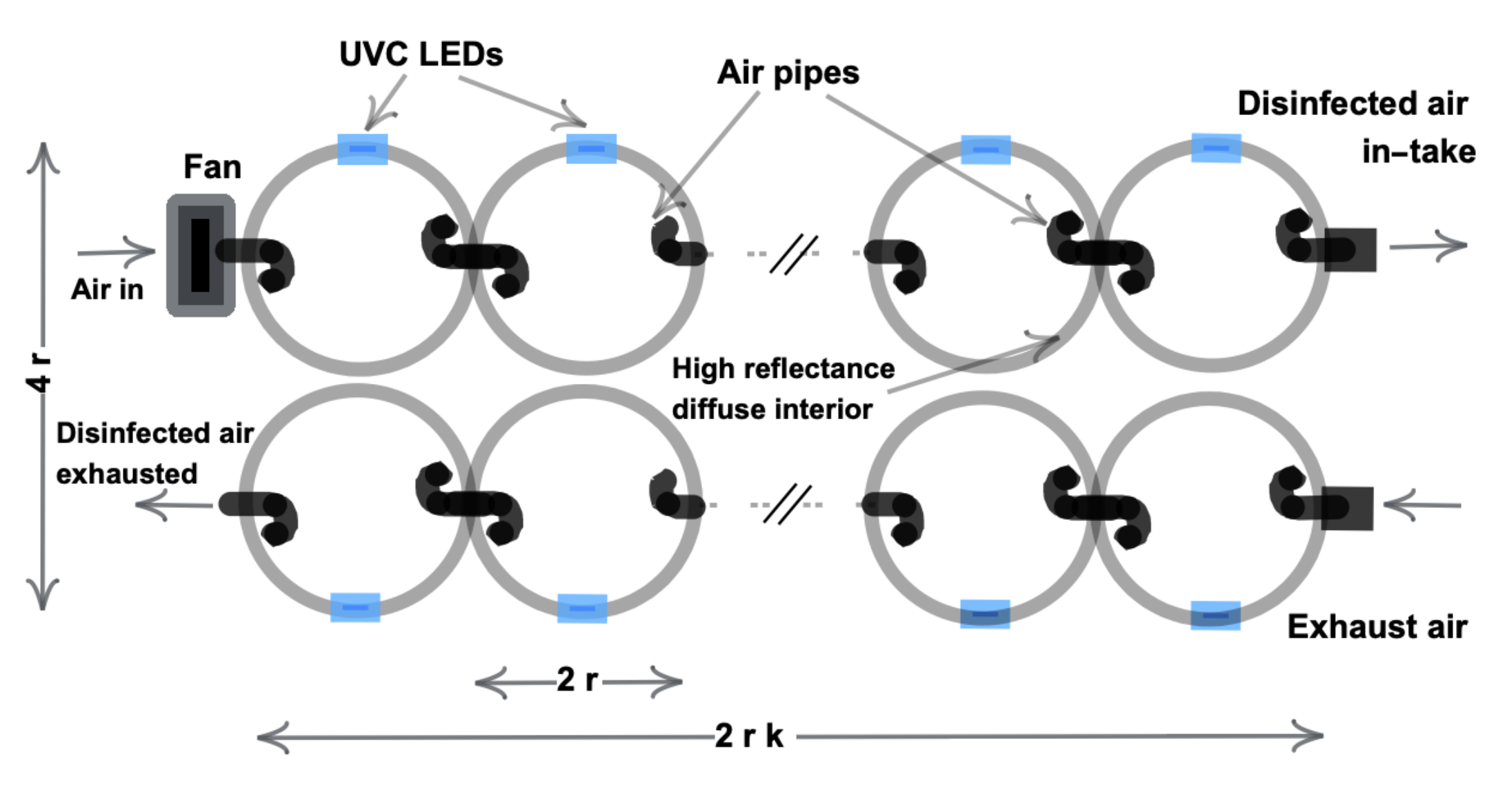}
      \caption{A schematic for the PUR$\diamond$GEM showing a series of $k$ spherical cavities in each disinfection direction. Each sphere has a radius $r$, leading to total linear dimension of $2r k=2 R$. A UV source (LED) in each sphere irradiates the internal cavities uniformly due to high reflectance interior walls. Air flow at required rate is propelled by example placement of fan as shown. The air is conducted through the series of cavities via curved pipes/buffered ports to prevent straight flow through. Drawing not to scale. See text for details. }
      \label{fig:diff}
   \end{centering}
\end{figure}

 A schematic of the PUR$\diamond$GEM is shown in Fig.~\ref{fig:diff}, configured for both in and out disinfection. The UV disinfection takes place in a series of $k$ spherical cavities, each of radius $r$ with high reflectance (diffuse) interiors.  A UV LED is mounted in each sphere, irradiating the entire sphere. There are several air inlets and outlets connecting the spheres and ultimately the user, configured to be furthest away from each other in each sphere to optimize diffuse turbulent air flow through the chamber~\cite{herbinet:hal-00880195, ayass2013mixing, crawford2014computational, esmaeelzade2019numerical}.   
 The input fan assembly may include a tight-weave cloth filter covering the opening to the chamber, and serves to act primarily as a particulate filter~\cite{konda2020aerosol}. More sophisticated particulate filters may be used if desired.
 A current sensor could be incorporated into the circuit design which alerts the user to current flow below a certain level, indicating LED/fan failure or low battery.

The quantity $D_0$ may be computed as follows:
 \be
 D_0 = \frac{\mathcal{E} ~P_{UV}}{SA_s} \frac{V_s}{V_{AF}} = \frac{P_{UV}}{3 V_F ({1-R_{eff}})} ~r \; ,
 \ee
where $P_{UV}$ is the UV source radiation in mW, $V_s=\frac{4}{3} \pi r^3$ is the volume of a sphere and $V_{AF}$ is the required air flow rate in cm$^3$/s. This quantity $D_0$ is what would be naively computed as the dose received by air flowing through. 

As discussed and validated experimentally in the previous section, the spherical shape of the UV disinfection chamber provides excellent spatial integration. Since there are minimal radiation hot or cold spots, the UV dose received by in-passing air volume becomes effectively independent of path taken through the sphere. However, travel time distributions, not average time, must be taken into account to accurately predict actual disinfection, and in general will reduce the actual pathogen elimination significantly independent of geometry. To compensate for this effect, we use a series of spheres which will effectively integrate the pathogen paths temporally, allowing for the achievement of much higher levels of actual disinfection.

The time air spends in the spherical cavities we consider will follow an exponential distribution distribution due to its turbulent motion, as has been verified both numerically and experimentally~\cite{herbinet:hal-00880195, ayass2013mixing, crawford2014computational, esmaeelzade2019numerical}. The turbulent nature of air flow was verified visually in a clear test sphere by filling it with incense smoke and then using a high speed phone camera to record the smoke movement as air was exhausted from the sphere using a fan at one of the ports.  For turbulent airflow, the probability that a differential volume element of air will spend a time $t$ as it flows through  a sphere is given by:
\be
\mathcal{P}(t) = \frac{1}{\tau} e^{-t/\tau}\;,
\ee
where $\tau = V_s/V_{AF}$. The expectation value for virus elimination in this volume of air is  computed by weighting it by this distribution. The usual convention of defining pathogen elimination in base 10 means one needs to convert
\be
\mathbb{S}=10^{-\frac{D}{D_{90}} } = e^{-\frac{D}{D_{90}} \ln 10}\;,
\ee
where $D_{90}$ is the dose required for 1-log (90\%) pathogen elimination, $D$ is dose received, and $\mathbb{S}$ is surviving fraction of pathogen. 
However, $D$ must be computed using the time that the pathogen actually spends being irradiated.
Therefore, the expectation value of the fraction of pathogen particles surviving passage through a sphere is given by
\be
\mathbb{S}_0=  \int_0^{\infty} e^{-\frac{\mathcal{E} P_{UV} \ln 10}{SA_s D_{90}} ~ t} ~ \mathcal{P}(t) ~ dt =  \frac{1}{1+ \frac{D_0}{D_{90}} \ln 10}\;.
\ee
Since the path of an air packet through a sphere is independent of its path through another sphere, given a series of $k$ identical spheres, each of radius $r$, irradiated by $P_{UV}$, pathogen elimination will be given by
\be
\mathbb{E}_k=1-\mathbb{S}_0^k\;.
\ee

 \begin{figure}[tb!]
   \begin{centering}
       \includegraphics[width = \textwidth, trim = 0cm 0cm 0.5cm .0cm, clip]{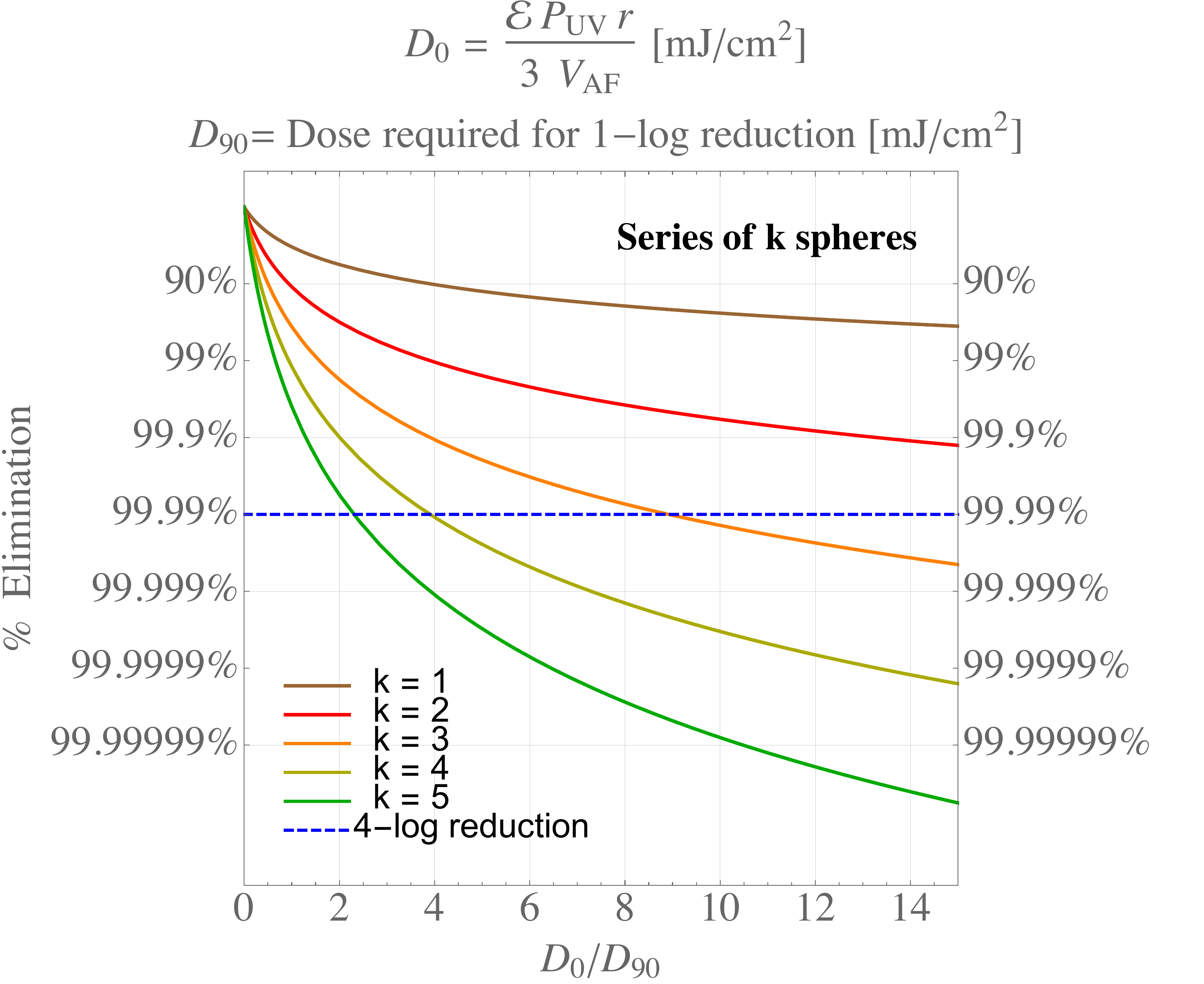}
      \caption{Pathogen elimination shown as a function of $D_0/D_{90}$ which encodes the enhancement factor, $P_{UV}$, radius $r$, the air volume flow rate $V_{AF}$, and the UV dose required for 1-log reduction $D_{90}$, for different number of spheres in series as denoted by the labels $k$. The dashed blue line corresponds to 4-log reduction in pathogen. }
      \label{fig:elim}
   \end{centering}
\end{figure}

In Fig.~\ref{fig:elim} we show the elimination possible in configurations with multiple spheres in series labeled by $k$ as a function of the dimensionless quantity $D_0/D_{90}$. The brown, red, orange, light green and dark green denote configurations with $k={1,2,3,4,5}$ spheres in series respectively. The blue dashed line denotes 4-log reduction or 99.99\% pathogen elimination. Increasing values of $D_0/D_{90}$ correspond to increase in enhancement factor, $P_{UV}$ or $r$, or a decrease in air flow rate $V_{AF}$. As can be seen, initially increasing $D_0/D_{90}$ there is significant increase in elimination, however, at a certain point for any number of spheres, the gain becomes progressively smaller. For the benchmark elimination of 4-log reduction, we see that one needs to practically have 3 or more spheres in series.

%**********************************************************************
\section{The PUR$\diamond$GEM Prototype } \label{sec:proto}
%**********************************************************************

The prototype constructed consisted of 2 spheres in each disinfection direction. The size of each sphere was dictated by the size requirement for installing the UVC LEDs manually. The UVC LEDs are only 3~mm~$\times$ 3~mm and are usually provided in tape form for easy machine application~\cite{LEDs}. However, the manufacturer provided the LEDs pre-mounted on 12~mm~$\times$12~mm PCB boards for prototype use. Therefore, 5 cm radius hemispheres with 1~cm diameter ports for pipes and 2~cm diameter ports for the LEDs were 3D printed using PLA. Pipes, fan grill/connector and exterior boxes for housing heat sink and wiring were similarly printed. Internal curved pipes, such as shown in schematic in Fig.~\ref{fig:diff}, were glued to the hemispheres. External pipes were sized to fit standard CPAP hoses~(22 mm outer diameter and 19 mm inner diameter). The interior of the hemispheres, including curved pipes, was then lined with four layers of PTFE. The LEDs were mounted on top of the heat sinks~\cite{heatsink} using thermal paste and adhesive, and attached to the base of the hemispheres using mechanical connectors providing an airtight seal. Heat pipes, which would have a significantly smaller fingerprint than conventional heat sinks, could be used if manufacturing commercially, however they are not easy to install manually. The LEDs were wired in series to a LED driver~\cite{driver} to maintain constant current of 500 mA through the circuit for optimal LED performance~\cite{LEDs}. Since the LED driver was enclosed in the case for the heat sinks and wiring, to prevent overheating it was also attached to a heat sink using thermal adhesive. The hemispheres were sealed together and a fan~\cite{fan} was attached to the input port of one of the spheres. The fan was wired in parallel to the LED circuit, and the entire circuit was  powered by a 12~V rechargeable battery~\cite{battery}. The weight of the device, excluding battery, is $\sim 1$ lb.  The whole contraption was then mounted to a water backpack~\cite{bp} via provided bungee cords (and extra components of backpack were removed). The wires were threaded through to front where battery was housed in a separately attached pocket to the backpack front straps.  

A freestanding GEM as constructed could be used for example for local air disinfection, or attached to intake and exhaust vents for ventilators. However, to construct a fully functional PUR$\diamond$GEM, a simple hood was constructed using a heavy duty plastic bag~\cite{bag}. Screw ports were 3D printed for attachment of  plastic hood to CPAP hoses which were then connected to the PUR$\diamond$GEM. Further, soft muslin cloth was attached via lacing to the bottom interior and the exterior of plastic hood for comfort.  A visor removed from lab googles~\cite{visor} was attached to the front for optical clarity. This also served the purpose of reducing problems with vision due to fogging of the interior. When in use, the whole bag is sealed around the neck using an elastic velcro strap. Inflation of the bag during use continuously confirms positive pressure is maintained. While there may be slight diffuse leakage of air outwards from the bottom, the air escaping in this way would have to go through multiple layers of cloth, effectively getting filtered.  

\begin{figure}[tb!]
   \begin{centering}
       \includegraphics[width = 0.29\textwidth, trim = 0cm 0cm 1cm 0cm, clip]{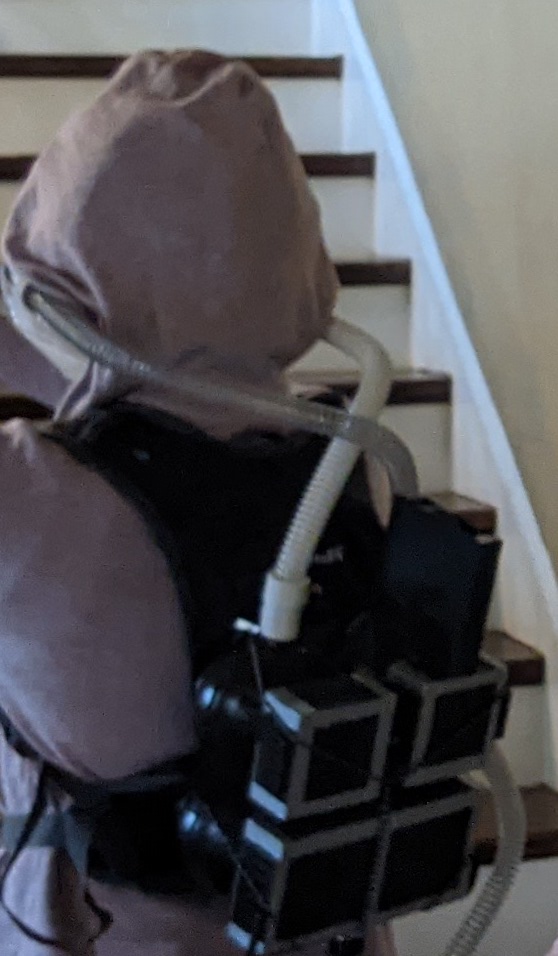}
       \hspace{.5in}
       \includegraphics[width = 0.41\textwidth, trim = 0cm 0cm 0cm 0cm, clip]{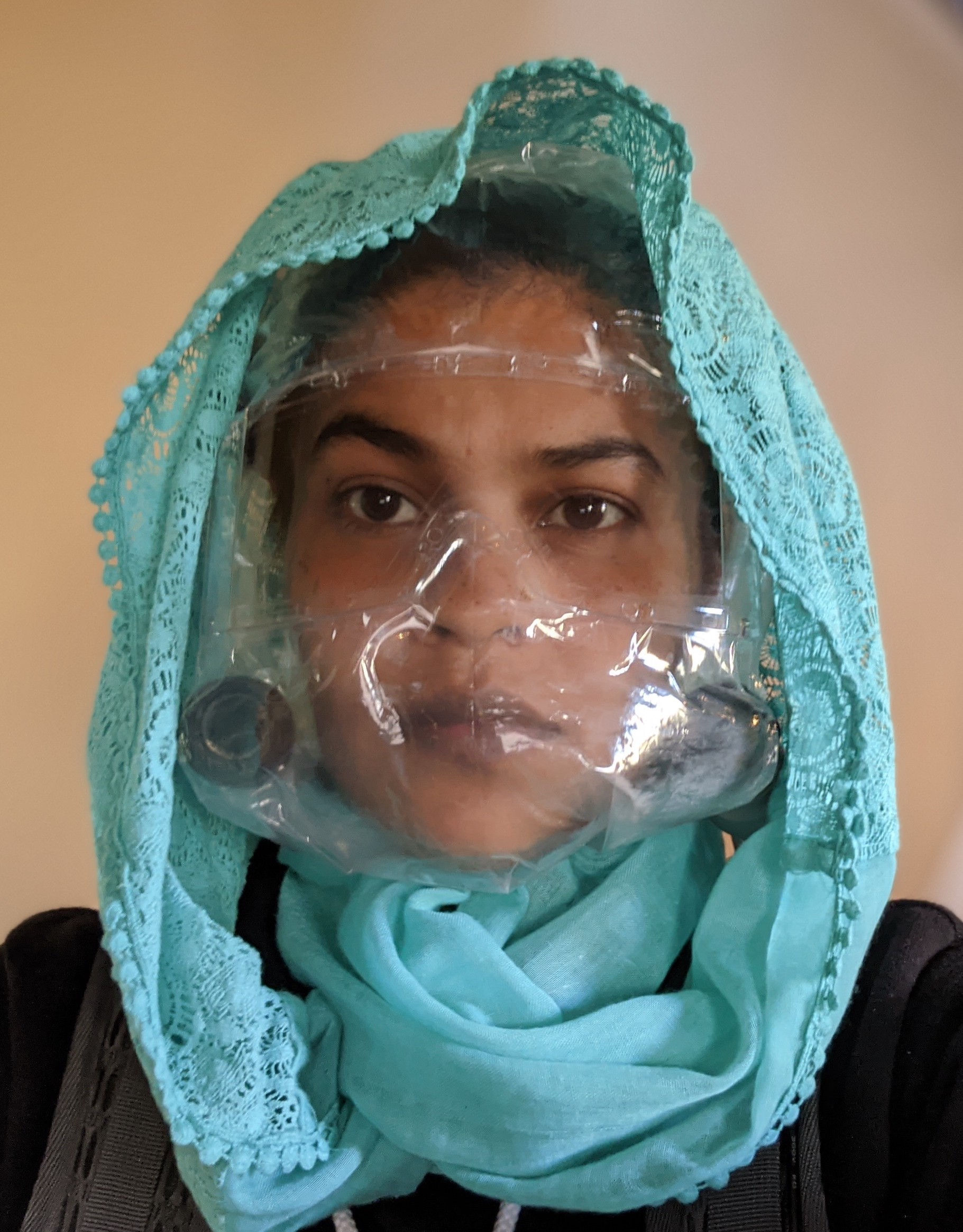}
      \caption{ The PUR$\diamond$GEM prototype as worn is shown:  {\it Left:} Back view showing device mounted on backpack and CPAP hoses connecting to the hood. {\it Right:} Front view of the hood.  Screw caps connecting plastic hood to CPAP hoses facilitating airflow can be seen. Inflated hood shows that positive pressure is maintained. }
      \label{fig:Proto}
   \end{centering}
\end{figure}

The finished PUR$\diamond$GEM prototype as worn is shown in Fig.~\ref{fig:Proto}. The back can be seen in the left panel and the front of hood is shown in the right panel.  Apart from the author, 3 adult and 1 child volunteer tested the complete functioning PUR$\diamond$GEM for comfort during use. The device was worn continuously for $\sim$ 2 hours at a time while performing normal activities. Airflow was measured~\cite{air} to be 30-35 L/min, and CO$_2$ measurements~\cite{co2} were $\lesssim$ 0.2\% after prolonged use.~\footnote{See Ref.~\cite{bunny} for validation of 20-30 L/min as being sufficient airflow for healthcare personnel.} The airflow was found to be sufficient for comfort, with minimal fogging. Further, in case of exertion leading to higher airflow requirement than what is propelled by the fan, due to the placement of the input air pipe, one can simply orally breath unrestricted fresh air from the pipe. However, it should be noted that disinfection for the duration of increased airflow will be reduced as shown in Fig.~\ref{fig:elim}.  To avoid heat build up, plastic bag should not be tucked in too tight around the head, such that airflow is maintained around head. There is a slight muffling/echoing of sound and some plastic noise crinkling when breathing. However, due to the flexible nature of the plastic, there is no real problem with communication. Clearly the PUR$\diamond$GEM prototype is somewhat bulky, however, due to its light weight, it doesn't hinder functionality in any significant way. Walking around, the impact on activity of wearing the PUR$\diamond$GEM is imperceptible, and for example if one is sitting down, the backpack can  easily be taken off and placed on the side.  

A note is in order for the required power for the fan to propel sufficient airflow and fan selection. While a myriad of fans in different combinations of size/power consumption/airflow/pressure are available for CPAP  machines for bulk ordering, it was not possible to obtain them for testing purposes. Therefore it was determined that easily available fans used for cooling in electronics would be used for the prototype construction.  Again a huge variety of fans is available, however,  air output is not geometrically optimized for input to a CPAP sized pipe. Multiple configurations for fan connectors to port air from fan to sphere input were 3D printed, but it was noted that as soon as any of the tested fans were connected to a different sized/shaped pipe, airflow  dramatically decreased. At the end, it was decided to simply use a direct short connector for the output of the fan to the sphere input. Certainly a more optimized fan/fan connector should be used which would reduce power consumption considerably if the PUR$\diamond$GEM is  manufactured commercially. It was further noted that there was effectively very little difference in airflow measured from fan connector outlet~(before being input to spheres) as compared to air outflow from the spheres, denoting a small pressure drop, as would be expected since the constriction to airflow is very short~(only immediately at the entrance and  exits to the spheres). However, once even a short CPAP hose is attached to the sphere outlet, a significant drop in airflow occurs. 

Summarizing, the constructed prototype  consisted of 2 sphere in each disinfection direction, with a radius of 5 cm each. For the constructed device, an enhancement factor of $\sim18$ was validated experimentally. 40 mW UV LEDs were used with an airflow of $\sim 30$ L/min. $D_{90}$ for SARS-COV2 should approximately be 1 mJ/cm$^2$~(see discussion and references in Ref.~\cite{PURGEM}), hence for current situation, the x-axis in Fig.~\ref{fig:elim} may be read in units of mJ/cm$^2$. This leads to 
\be
\mathbb{E}_2=1-\mathbb{S}_0^2 =  1-\left[ \frac{1}{1+ \frac{18\times 40~\mathrm{mW} \times 5 \mathrm{cm}}{3\times 500 \mathrm{cm^3/s} \times1~ \mathrm{mJ/cm^2}} \ln 10}\right]^2 = 1-\left[ \frac{1}{1+ 2.4\ln 10}\right]^2 = 97.65\%\;.
\ee

Significantly smaller spheres for example with $r=2$ cm~(ping-pong ball size) may be manufactured commercially. In the lab prototype, there was probably some degradation in the enhancement factor due to manual application of the PTFE wrap to the interior. Commercial manufacturing could be expected to minimize such defects. Using the same air flow rate, but higher power LEDs~(which are available) a series of 2 spheres would give similar disinfection as the prototype
\be
\mathbb{E}_2=1-\mathbb{S}_0^2 =  1-\left[ \frac{1}{1+ \frac{20\times 80~\mathrm{mW} \times 2 \mathrm{cm}}{3\times 500 \mathrm{cm^3/s} \times1~ \mathrm{mJ/cm^2}} \ln 10}\right]^2 = 1-\left[ \frac{1}{1+ 2.13\ln 10}\right]^2 = 97.14\%\;,
\ee   
which is probably sufficient for commercial use. For higher disinfection, keeping the same form factor, 
5 such spheres would give
\be
\mathbb{E}_5=1-\mathbb{S}_0^5 =  1-\left[ \frac{1}{1+ \frac{20\times 80~\mathrm{mW} \times 2 \mathrm{cm}}{3\times 500 \mathrm{cm^3/s} \times1~ \mathrm{mJ/cm^2}} \ln 10}\right]^5 = 1-\left[ \frac{1}{1+ 2.13\ln 10}\right]^5 = 99.99\%\;,
\ee   
which may be relevant for health care use. 

Finally we note that due to their light weight, a series of such small footprint spheres could be configured to be attached to the sides of helmet/hood or around neck eliminating the need for a backpack.  If the air input/output from the spheres is directly connected to the helmet/hood, the need for CPAP hoses is eliminated, further reducing footprint and power consumption of fans. The rechargeable battery powering such a one-piece PUR$\diamond$GEM helmet/hood could be clipped to belt or placed in a pocket.

%*********************************************************
\acknowledgments
%*********************************************************
I thank T. Sakamoto and C. Kelly for useful discussions and the use of lab equipment. I am also grateful to Kamran Shah, Iman Shah, Nahid Raees and  Vajahat Raees for spending many hours wearing the PUR$\diamond$GEM for testing purposes.

\bibliographystyle{JHEP.bst}
\bibliography{theBib}

\end{document}